\documentclass{article}

\usepackage{arxiv}

\usepackage[utf8]{inputenc} % allow utf-8 input
\usepackage[T1]{fontenc}    % use 8-bit T1 fonts
\usepackage{hyperref}       % hyperlinks
\usepackage{url}            % simple URL typesetting
\usepackage{booktabs}       % professional-quality tables
\usepackage{amsfonts}       % blackboard math symbols
\usepackage{nicefrac}       % compact symbols for 1/2, etc.
\usepackage{microtype}      % microtypography

\usepackage{amsmath,amssymb}
\usepackage{graphicx}
\usepackage{textcomp}
\usepackage{xcolor}

\usepackage{mathtools}

\usepackage{algorithm,algpseudocode}
\makeatletter
\newcommand{\algmargin}{\the\ALG@thistlm}
\makeatother
\algnewcommand{\parState}[1]{\State%
    \parbox[t]{\dimexpr\linewidth-\algmargin}{\strut\hangindent=\algorithmicindent \hangafter=1 #1\strut}}
    
\DeclarePairedDelimiter\abs{\lvert}{\rvert}%
\DeclarePairedDelimiter\norm{\lVert}{\rVert}%

\makeatletter
\let\oldabs\abs
\def\abs{\@ifstar{\oldabs}{\oldabs*}}
\let\oldnorm\norm
\def\norm{\@ifstar{\oldnorm}{\oldnorm*}}
\makeatother

\title{Sound Event Recognition in a Smart City Surveillance Context}

\author{
  Tito Spadini\thanks{https://spadini.info} \\
  Laboratório de Sinais e Sistemas\\
  Universidade Federal do ABC\\
  Santo André, SP \\
  \texttt{tito.caco@ufabc.edu.br} \\
   \And
   Dimitri Leandro de Oliveira Silva \\
  Laboratório de Sinais e Sistemas\\
  Universidade Federal do ABC\\
  Santo André, SP \\
  \texttt{dimitri.leandro@gmail.com} \\
   \And
 Ricardo Suyama \\
  Laboratório de Sinais e Sistemas\\
  Universidade Federal do ABC\\
  Santo André, SP \\
  \texttt{ricardo.suyama@ufabc.edu.br} \\
}

\begin{document}
\maketitle

\begin{abstract}
Due to the growing demand for improving surveillance capabilities in smart cities, systems need to be developed to provide better monitoring capabilities to competent authorities, agencies responsible for strategic resource management, and emergency call centers. This work assumes that, as a complementary monitoring solution, the use of a system capable of detecting the occurrence of sound events, performing the Sound Events Recognition (SER) task, is highly convenient. In order to contribute to the classification of such events, this paper explored several classifiers over the SESA dataset, composed of audios of three hazard classes (gunshots, explosions, and sirens) and a class of casual sounds that could be misinterpreted as some of the other sounds. The best result was obtained by SGD, with an accuracy of 72.13\% with 6.81 ms classification time, reinforcing the viability of such an approach.
\end{abstract}

% keywords can be removed

%   ========================
%   ----- INTRODUCTION -----
%   ========================
\section{Introduction}
\label{sec:introduction}

Among the many concerns typically addressed when it comes to smart cities are subjects associated with public security issues, which, given the circumstances, eventually involves the use of low-cost sensor-based surveillance capabilities strategically positioned to create a virtual monitoring layer.

Such remote monitoring could be done in the usual way, which relies only on camcorders. However, in order to cover the entire region of interest, large numbers of camcorders would be required to record the environment continuously. Difficulties related to this approach include the high costs of equipment and infrastructure demanded for implementation, the difficulty of maintenance, the high energy cost of the process, the further degradation with weathering, the increased difficulty of stealth installation, the high demand for processing power, the reliance on huge storage capacities, and, perhaps one of the worst among those listed here, the reliance on human guards who will need to watch the images as closely as possible.

A more viable alternative to work around such problems may be to use ambient sound; thus, instead of filming the locations, audios of the environments being monitored will be recorded. Instead of using people who would listen to the environment all the time waiting for a possible occurrence, it would be more convenient to build a system that automatically detects potential public safety events and takes action corresponding to the type of problem detected.

Despite the existence of numerous works that manifest the feasibility of using sound as surveillance resources \cite{4425280, 1521669, 1521503, 1540194, 4959546}, there are still few studies showing concern about cost, scalability, and practicality. In general, the most recent works tend to perform such tasks on supercomputers equipped with powerful GPUs and use costly Deep Learning algorithms with deep and complex neural network architectures \cite{8751822, 7003973, 8737801}. Accuracy is probably the most widely used metric for assessing classifier performance. However, in the context of smart cities, it is also convenient to evaluate classifier processing time for classification to be performed because less costly algorithms tend to be desirable for embedded hardware deployments.

This paper is organized as follows: section \ref{sec:dataset} will explain the composition of the dataset used in this paper; the section \ref{sec:methodology} will list features and classifiers, and explain the classification methodology; the section \ref{sec:results} will display the results; the section \ref{sec:discussion} will bring the discussions; and the section \ref{sec:conclusion} will bring the conclusions.

%   ===================
%   ----- DATASET -----
%   ===================
\section{Dataset}
\label{sec:dataset}

The Sound Events for Surveillance Applications (SESA) dataset \cite{tito_spadini_2019_3519845} is comprised of three unbalanced classes that represent security threats (gunshot, explosion, and siren) and a casual class that encompasses a wide range of audios that could be misclassified as any of the security risk classes. The files are separated into training and test folders. All dataset entries follow the same configuration: WAV, mono, 16 kHz, and 8-bit lasting up to 33 seconds each.

The ``gunshot'' class has firearm shooting of various types and calibers, considering different scenarios where the shooting was made. The ``explosion'' class encompasses explosions of small artifacts such as hand grenades and small explosives, as well as accidental explosions of homes and commercial buildings. The ``siren'' class, despite its name, encompasses both sirens, car alarms, and building alarms. Moreover, the ``casual'' class is made up of several audio samples that have many similarities to at least one of the other three classes, such as firework sounds, thunderstorms, hammering, and many others.

\setlength{\tabcolsep}{10pt}
\begin{table}[ht]
    \centering
    \caption{Features explored in this paper.}
    \begin{tabular}{llc}
        \toprule
            \textbf{Feature}        & \textbf{Position} & \textbf{Reference} \\
        \midrule
            \rule{0pt}{2.0ex}
            Chroma CENS             & [1--12]           & \cite{ChromaCENS} \\
            \rule{0pt}{2.0ex}
            Constant-Q Chromagram   & [13--24]          & \cite{ChromaCQT_1,ChromaCQT_2} \\
            \rule{0pt}{2.0ex}
            Chromagram              & [25--36]          & \cite{Chromagram} \\
            \rule{0pt}{2.0ex}
            Melspectrogram          & [37--56]          & \cite{Melspectrogram} \\
            \rule{0pt}{2.0ex}
            MFCC Slaney             & [57--76]          & \cite{MFCCSlaney} \\
            \rule{0pt}{2.0ex}
            MFCC HTK                & [77--96]          & \cite{MFCCHTK} \\
            \rule{0pt}{2.0ex}
            MFCC Delta              & [97--116]        & \cite{MFCCDeltas} \\
            \rule{0pt}{2.0ex}
            MFCC Delta-Delta        & [117-136]        & \cite{MFCCDeltas} \\
            \rule{0pt}{2.0ex}
            Root-Mean-Square        & [137]              & \cite{schuller2013intelligent} \\
            \rule{0pt}{2.0ex}
            Spectral Centroid       & [138]              & \cite{eyben2015real} \\
            \rule{0pt}{2.0ex}
            Spectral Bandwidth      & [139]              & \cite{SpecBandwidth} \\
            \rule{0pt}{2.0ex}
            Spectral Contrast       & [140-146]        & \cite{SpecContrast} \\
            \rule{0pt}{2.0ex}
            Spectral Flatness       & [147]             & \cite{eyben2015real} \\
            \rule{0pt}{2.0ex}
            Spectral Rolloff        & [148]             & \cite{eyben2015real} \\
            \rule{0pt}{2.0ex}
            Zero-Crossing Rate      & [149]             & \cite{schuller2013intelligent} \\
        \bottomrule
    \end{tabular}
    \label{tab:features}
\end{table}

In order to accurately classify sound events, characteristics are required that enable classifiers to notice patterns in features according to each category. Therefore, there must be features that provide enough information for this; and not all features contribute positively to this process. Thus, a large number of temporal, frequency and cepstral features are extracted so that, after being evaluated, it can proceed to the next steps in the pipeline. The features considered in this paper are listed in Table \ref{tab:features}. For feature extraction, a 200 ms window and a 50\% overlap were adopted.

%   =======================
%   ----- METHODOLOGY -----
%   =======================
\section{Methodology}
\label{sec:methodology}

\begin{figure}[ht]
    \centering
    \includegraphics[width=0.75\textwidth]{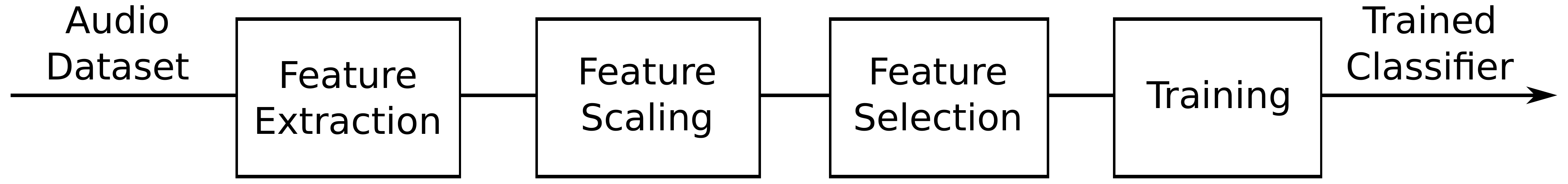}
    \caption{Model training process pipeline.}
    \label{fig:pipeline}
\end{figure}

After the feature extraction process, data was preprocessed to adjust the values of all features so that the values respect the same range; Then, a feature selection was made to reduce the dimensionality of the problem using first a filter that eliminates features with lower variance and then the application of PCA to reshape the geometry of the problem, reducing feature hyperspace. Then, models were trained and used to classify the test samples, considering a 3-fold for cross validation. Accuracy and time required to classify were obtained to evaluate models. The Fig. \ref{fig:pipeline} shows the pipeline used to train the classifiers.

%   ===================
%   ----- RESULTS -----
%   ===================
\section{Results}
\label{sec:results}

After being trained, each model was used to classify the test samples, so the accuracy and time required to classify were obtained, as shown in Table \ref{tab:results}.

\setlength{\tabcolsep}{10pt}
\begin{table}[ht]
    \centering
    \caption{Results of accuracy and time required to classify for each classifier.}
    \begin{tabular}{lcc}
        \toprule
            \textbf{Classifier} & \textbf{Time [ms]}        & \textbf{Accuracy [\%]} \\
        \midrule
            \rule{0pt}{2.0ex}
            AdaBoost            & $5453.48 \pm 238.46$      & $65.46 \pm 3.00$ \\
            \rule{0pt}{2.0ex}
            Bagging             & $5158.75 \pm 224.08$      & $\mathbf{73.05 \pm 1.14}$ \\
            \rule{0pt}{2.0ex}
            Decision Tree       & $88.73 \pm 5.30$          & $65.38 \pm 1.18$ \\
            \rule{0pt}{2.0ex}
            Gradient Boosting   & $1027.63 \pm 34.95$       & $71.88 \pm 0.87$ \\
            \rule{0pt}{2.0ex}
            KNN                 & $21546.18 \pm 7265.36$    & $70.12 \pm 0.38$ \\
            \rule{0pt}{2.0ex}
            Perceptron          & $6.93 \pm 0.26$           & $68.20 \pm 2.81$ \\
            \rule{0pt}{2.0ex}
            Passive Aggressive  & $6.97 \pm 0.74$           & $69.40 \pm 3.25$ \\
            \rule{0pt}{2.0ex}
            Random Forest       & $319.17 \pm 34.19$        & $70.96 \pm 1.52$ \\
            \rule{0pt}{2.0ex}
            Ridge               & $7.16 \pm 0.99$           & $69.51 \pm 13.70$ \\
            \rule{0pt}{2.0ex}
            SGD                 & $\mathbf{6.81 \pm 0.16}$  & $72.13 \pm 2.78$ \\
            \rule{0pt}{2.0ex}
            SVM                 & $6093.75 \pm 1978.99$     & $71.75 \pm 1.72$ \\
        \bottomrule
    \end{tabular}
    \label{tab:results}
\end{table}

Based on Table \ref{tab:results}, it is noticeable that the fastest algorithm for processing the classification task is the SGD, which was able to perform the classification in 6.81 milliseconds; and the algorithm that exhibited the highest average accuracy is Bagging, which reached 73.05\%.

%   ======================
%   ----- DISCUSSION -----
%   ======================
\section{Discussion}
\label{sec:discussion}

Although not as fast as SGD, Perceptron, Passive Aggressive, and Ridge also drew attention to their speed; Like SGD itself, Gradient Boosting and SVM achieved high accuracy. Still, the algorithm that exhibited the best overall performance was SGD, as it obtained the second-highest accuracy among the algorithms used and was the fastest one.

On the other hand, some of the algorithms drew attention for negative reasons, such as KNN, which took an average of more than 21 seconds to classify, which is more than 3000 times the time required by the fastest algorithm. Furthermore, the Ridge algorithm also exhibited an unexpectedly negative result in its accuracy, with a substantially high error compared to the other classifiers.

%   ======================
%   ----- CONCLUSION -----
%   ======================
\section{Conclusion}
\label{sec:conclusion}

Based on the results obtained, it can be concluded that the implementation of a sound event recognition system is feasible and that there are already algorithms capable of performing such a class classification task for a smart city surveillance context. A priori, due to its performance, it may be convenient to exploit the SGD classifier in embedded hardware implementations. However, some of the other algorithms also exhibited performance close to SGD, indicating that it may be convenient to do more studies to explore them in more detail. Besides that, a more careful analysis of dimensionality reduction and feature selection processes can yield significant performance increases.

\section*{Acknowledgment}

The authors are grateful for the support received from CAPES and from the Conselho Nacional de Desenvolvimento Científico e Tecnológico -- CNPq.

\bibliographystyle{unsrt}  
\bibliography{main.bib}  %%% Remove comment to use the external .bib file (using bibtex).


% George Kour and Raid Saabne.
% \newblock Real-time segmentation of on-line handwritten arabic script.
% \newblock In {\em Frontiers in Handwriting Recognition (ICFHR), 2014 14th
%   International Conference on}, pages 417--422. IEEE, 2014.

% George Kour and Raid Saabne.
% \newblock Fast classification of handwritten on-line arabic characters.
% \newblock In {\em Soft Computing and Pattern Recognition (SoCPaR), 2014 6th
%   International Conference of}, pages 312--318. IEEE, 2014.

% \bibitem{hadash2018estimate}
% Guy Hadash, Einat Kermany, Boaz Carmeli, Ofer Lavi, George Kour, and Alon
% \newblock Estimate and replace: A novel approach to integrating deep neural
%   networks with existing applications.
% \newblock {\em arXiv preprint arXiv:1804.09028}, 2018.

% \end{thebibliography}

\end{document}